\documentclass[twocolumn,showpacs,preprintnumbers,amsmath,amssymb,prl]{revtex4}

\usepackage{graphicx}
\usepackage{dcolumn}
\usepackage{bm}

\begin{document}

\title{Binding between endohedral Na atoms in Si clathrate I;\\
 a first principles study}

\author{Hidekazu Tomono}
 \affiliation{Department of Mechanical
    Engineering Informatics, School of Science and
    Technology, Meiji University, Kawasaki 214-8571 Japan}

\author{Haruki Eguchi}
    \affiliation{Department of Mechanical
     Engineering Informatics, School of Science and
     Technology, Meiji University, Kawasaki 214-8571 Japan}
     \affiliation{Advanced Applied Science Department, 
     Research Laboratory, IHI Corporation, Yokohama 235-8501
     Japan}

\author{Kazuo Tsumuraya}
 \email{abinitio[atmark]isc.meiji.ac.jp}
 \affiliation{Department of Mechanical
    Engineering Informatics, School of Science and
    Technology, Meiji University, Kawasaki 214-8571 Japan}

\begin{abstract}
We investigate the binding nature of the endohedral sodium atoms with
the density functional theory methods, presuming that the clathrate I
consists of a sheaf of one-dimensional connections of Na@Si$_{24}$ cages 
interleaved in three perpendicular directions.
Each sodium atom loses 30\% of the 3$s^1$ charge to the frame, forming
an ionic bond with the cage atoms;
the rest of the electron contributes to the covalent bond between the
nearest Na atoms.
The presumption is proved to be valid; the configuration of the two Na
atoms in the nearest Si$_{24}$ cages is more stable by $0.189$ eV than
that in the Si$_{20}$ and Si$_{24}$ cages. 
The energy of the beads of the two distorted Na atoms is more stable
by $0.104$ eV than that of the two infinitely separated Na atoms.
The covalent bond explains both the preferential occupancies in the
Si$_{24}$ cages and the low anisotropic displacement parameters of
the endohedral atoms in the Si$_{24}$ cages in the [100] directions
of the clathrate I.
\end{abstract}


\maketitle

\section{Introduction}
The understanding of the mechanism of cohesion of condensed matter is
essential for solid state physics. 
Silicon clathrates are compounds with endohedral atoms
in the cages of the host frame network and expanded phases of 
diamond type silicon crystal.
Cros {\it et al}~\cite{bib:Cros1965}.\ inspired by the structure
of the clathrate natural gas hydrates, have first synthesized silicon
clathrate I containing Na atoms.
Group 14 clathrates I have been successively synthesized only when
alkaline~\cite{bib:Bobev2000,bib:Gryko1996,bib:Nolas2001} or alkaline
earth metal atoms~\cite{bib:Cordier1991} or Cl, Br, or I in group 17
atoms~\cite{bib:Reny2000} are encapsulated into the clathrate cages.
The electro-negativity differences between these host and guest
atoms are smaller than those in the ionic crystals.
If the host and guest atoms have large differences, then the induced
electron transfer forms the ionic compounds with simple structures like
NaCl or CsCl type structures.

To date we have found few reports on the role of the endohedral atoms
in the cohesion of the group 14 clathrates.
The electron charge transfers, from the endohedral Na atom to the frame
silicon atoms, have been predicted in clathrates
I~\cite{bib:Zhao1999,bib:Gatti2003} and a partial transfer in a
Ba@Si$_{20}$ cluster~\cite{bib:Nagano2001}.
In clathrate II, a displacement of the guest atoms has been predicted
to be $0.17$ \AA \ from the center of the Si$_{28}$ cage and been 
explained the displacement to be due to a combination of the Jahn-Teller
and Mott transition~\cite{bib:Demkov1994}.
Brunet {\it et al}~\cite{bib:Brunet2000}.\ have observed the displacement
using EXAFS (extended x-ray absorption fine structure) analysis.
The Na atom was displaced away from the Si$_{28}$ cage-center
toward the center of a hexagonal ring by
$0.9 \pm 0.02$ \AA ~\cite{bib:Brunet2000}.
Libotte {\it et al}~\cite{bib:Libotte2003}.\ have calculated the
displacements of the endohedral Na atoms in the clathrate II and found
the displacements to be $0.456$ \AA \ from the {\it ab initio}
calculation and $0.91$ \AA \ from a tight-binding calculation. 
Tournus {\it et al}.\ have observed the displacements to be $1$ \AA \
in the the Si$_{28}$ cage of the clathrate II Na$_{2}$@Si$_{34}$ and
$2$ \AA \ in clathrate II Na$_{6}$@Si$_{34}$~\cite{bib:Tournus2004}.
They also calculated the displacements of the Na atoms in the
Si$_{28}$ cage as $0.65$ \AA \ from the supercell calculation of
the Na$_{2}$@Si$_{50}$H$_{44}$ cluster with the periodic 
DFT calculation. 
They proposed a possibility of the displacements to be due to the
Peierls or Jahn-Teller effect.

Recently one of the authors has reported the displacements of the Na
atoms in the two adjacent Si$_{28}$ cages hydrogenated to terminate
the dangling bonds of the Si atoms on the surface of the
clusters~\cite{bib:Takenaka2006}. 
Each Na atom displaced by $0.63$ \AA \ away from their centers
of the cages to form a dimer between the endohedral Na atoms.
The displacements was attributed to the formation of covalent bond
between the endohedral Na atoms. 
They also found the electron charge transfered from the endohedral
atoms to the silicon atoms. 

So the following questions arise: 
What is the binding between the endohedral atoms in the cohesion of the
clathrates?
Why do not the host-guest combinations crystallize into the simple ionic
structures? 
In the following we use a first principles analysis to address the
questions through investigating the guest-guest and the host-guest
interactions in the clathrate I\@. 

\begin{figure}
\begin{center}
\includegraphics{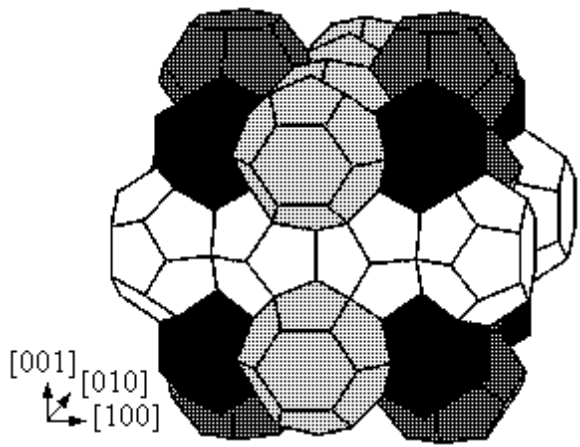}
\end{center}
\caption{\label{fig:schematic}
Polyhedron structure of the clathrate I Si$_{46}$ by extending the
simple cubic unit cell.
Two horizontal white bamboos of the polyhedron are a one-dimensional
bamboo like connection of the tetrakaidecahedron (Si$_{24}$) cages
in the [100] direction.
The connections, arranged in three perpendicular directions with spacing 
$a$ of lattice constant, forms the black voids of the pentagonal
dodecahedron.
This structure is the clathrate I Si$_{46}$, free of endohedral atoms,
consisting of the tetrakaidecahedra only.}
\end{figure}

Fig.\ \ref{fig:schematic} shows a schematic drawing of the polyhedral
structure of the clathrate I\@.
The structure is special in that it consists of the bamboo like
Si$_{24}$ cages only;
the cages (white) are arranged in bamboos, with spacing $a$ of lattice
constant, in one-dimensional horizontal direction sharing hexagonal
rings as the bamboo joints between the adjacent Si$_{24}$ cages.
Weaving the bamboos in three dimensions with common pentagonal surfaces
forms voids shown with black polyhedron regions in fig.\ \ref{fig:schematic}.
Each void is a pentagonal dodecahedron, separated in space, 
located at bcc position with a different orientation. 
All the previous papers have identified the existence of the Si$_{20}$
cages in the clathrate I\@. 
However we presume the structure to consist of only the Si$_{24}$ cages;
fig.\ \ref{fig:schematic} shows that Si$_{20}$ cages are merely accidental voids
in the weaved bamboos in the three dimensions.
The voids just correspond to $\alpha $ cages in zeolites, although the
voids in the clathrate I are far smaller than the ones in the zeolites. 
The accidental voids are predicted to have a minor role in the 
cohesion of the clathrate I\@. 
Although this view on the clathrate I structure has been
neglected so far, the experimental preferential 
occupancies of the endohedral atoms in the Si$_{24}$
cages~\cite{bib:Cros2737,bib:Yamanaka103,bib:Ramanchandran626}
and the experimental anisotropic displacement parameters
support this bamboo model.

So presuming the clathrate I as consisting of the bamboo structures
in the three perpendicular directions, we analyze the bonding nature
between the endohedral atoms in the clathrate.
First, we calculate the relaxed geometries of the one-dimensional
clusters with different numbers of the Si$_{24}$ cages using real-space
DFT method and show the binding nature between the guest atoms;
the dimer formation due to the covalent bonding between the adjacent
endohedral Na atoms and the charge transfer from the Na atoms to
the cage atoms. 
Next, we evaluate the cohesion energy of the chained Na atoms
using the periodic DFT method.

\section{Computational details}
We perform the real-space DFT calculation for the
hydrogenated bamboo structures using the generalized gradient approximation
of Perdew, Burke and Ernzerhof (GGA-PBE)~\cite{bib:GGA-PBE}.
We use frozen core 1$s^2$2$s^2$2$p^6$
approximation for the Na and the silicon atoms and the atomic
orbitals with valence 3$s^1$ orbital for the Na atom and the
valence 3$s^2$3$p^2$ orbitals for the silicon atom
each for which we use double atomic functions for each orbital. 
No smearing for occupations is applied to the final geometrical optimization.
Since we regard the bamboo clusters as representing the essential
aspects of the clathrate I, we add hydrogen atoms to the three
coordinated silicon atoms on the surface of the bamboo structures,
to mimic both the electronic density of states (DOS) and the bonding
configurations in the clathrate I with the clusters. 
The hydration on the surface of clusters mimics almost the same electronic
states as in the crystalline clathrates.
The calculated displacements~\cite{bib:Tournus2004,bib:Takenaka2006}
in the hydrogenated double Si$_{28}$ cages in the clathrate II
coincided with not only the experimentally observed displacements
$0.9$ \AA ~\cite{bib:Brunet2000} or $1$ \AA ~\cite{bib:Tournus2004} 
but also the calculated displacements $0.456$ \AA ~\cite{bib:Libotte2003}
or $0.91$ \AA ~\cite{bib:Libotte2003} in the crystalline clathrates II\@.
This hydration has enabled the states of the dangling bonds on the
surface of the bamboo structure to shift lower side in energy
as will be shown in fig.\ \ref{fig:dos}. This hydration has realized
the same features as in the DOS's of the clathrate
Ba$_{8}$@Si$_{46}$~\cite{bib:Nakamura2005}.
We use an ADF code~\cite{bib:ADF1,bib:ADF2}, which uses a linear
combination of Slater type orbitals.
To evaluate the cohesion energy of the chain of the two endohedral
Na atoms in the clathrate I, we use a periodic DFT code 
PHASE~\cite{bib:PHASE} with the norm conserving pseudopotentials
for the Na and the Si atoms.
For the periodic DFT calculations, Brillouin zones are sampled at
$\Gamma $ and $\mathbf{X}$ point set. 
Markov, Shah and Payne have shown that this set is an efficient
$\mathbf{k}$-point set to remove defect interactions in the periodic
cells~\cite{bib:Makov1996}.
The numbers of planewaves are kept 13,805 at $\Gamma $ point and
16,184 at $\mathbf{X}$ for any lattice constant.
It corresponds to set the cutoff energy to be $20.0$ Ry ($272.11$ eV)
at 11.0 \AA .
We use the PBE~\cite{bib:GGA-PBE} exchange and correlation functionals
for the electron correlations for the periodic DFT calculations.
We use the spin unrestricted calculations for both the real-space 
and the periodic calculations with the convergence of interatomic
forces reduced below to within $9.45\times 10^{-3}$ H/\AA \ 
($5.0 \times 10^{-3}$ H/bohr).

\section{Results}
\begin{figure}
\begin{center}
\includegraphics{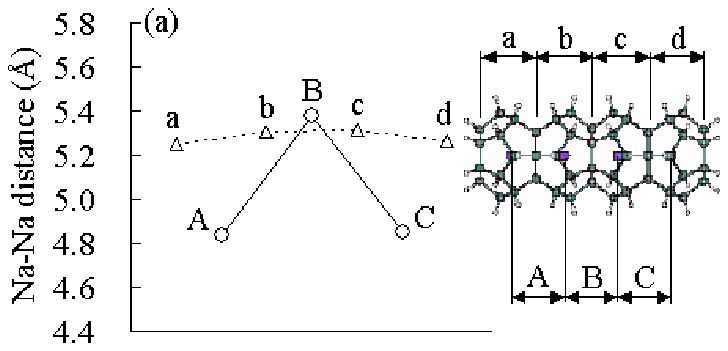}
\includegraphics{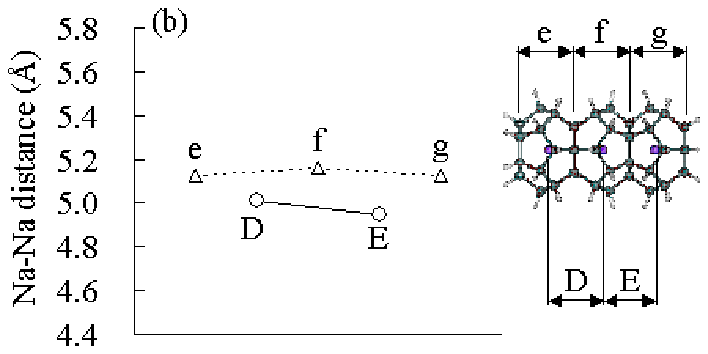}
\end{center}
\caption{\label{fig:distance} The inter-Na distances of
(a) four-caged Na$_4$@Si$_{78}$H$_{60}$ bamboo cluster and
(b) three-caged Na$_3$@Si$_{60}$H$_{48}$ cluster,
where the triangles are the inter-hexagonal
distances in the bamboo structures. 
The lines are for visual guidance.}
\end{figure}

The distances between the endohedral Na atoms in the relaxed four-caged
bamboo structure Si$_{78}$H$_{60}$ are shown in fig.\ \ref{fig:distance}(a)
as an example of the relaxed structures of even 
%
\begin{figure}
\begin{center}
\includegraphics{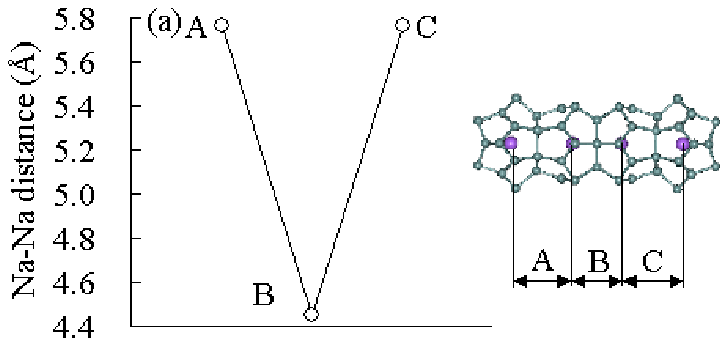}
\includegraphics{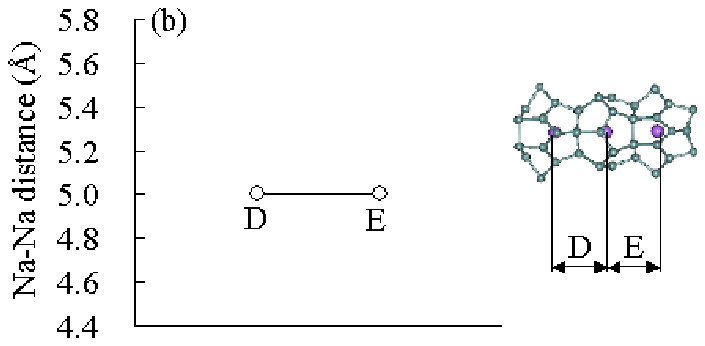}
\end{center}
\caption{\label{fig:distanceHfree} The inter-Na distances of
(a) four-caged Na$_4$@Si$_{78}$ bamboo cluster and
(b) three-caged Na$_3$@Si$_{60}$ cluster.
The line is for visual guidance.}
\end{figure}
number of cage clusters. Although the distances between the hexagonal
rings are almost constant, the inter-Na distances A and C are however
shorter than the inter-hexagonal distances:
the inter-Na distances A (4.84 \AA ) and C (4.85 \AA ) at the ends of
the bamboo structure are shorter than the distance B (5.38 \AA ). 
The short distances are induced by a bonding between the Na atoms.
The sum of the shorter and the longer distances is 
$10.21 \sim 10.23$ \AA \ which almost equals the experimental lattice
constant $10.19 \pm 0.02$ \AA \ of clathrate I
Na$_8$@Si$_{46}$~\cite{bib:Cros1971}.
We show the distances in the three-caged Na$_3$@Si$_{60}$H$_{48}$
cluster in fig.\ \ref{fig:distance}(b) as an example of odd number
of the cages.
The inter-Na distances, which are smaller than the ones between
the adjacent hexagonal rings, are almost the same for each endohedral
atom. 
A balance of forces exists between the central Na atom and the 
adjacent two Na atoms.
So the small inter-Na distances A and C in fig.\ \ref{fig:distance}(a)
are induced by the dimer formation between the Na atoms.
The formation may leads to a Peierls distortion in the bamboo
clusters.

For the Peierls distortion of one-dimensional case with a free
boundary condition, the inter-atom distances at the edge are different
from those in a periodic boundary condition.
Since two neighbor atoms at the edges form an edge state in
Peierls gap~\cite{bib:Figge2002}, their distances are longer than the
ones of inner inter-atom bonding since they are located at the free
boundary edge.
The present bamboo structures have the free boundary condition.
Thus the Si-H bonds at the edges do form longer Si-H bond distances.
The Na atoms just inside the bonds in the four caged structure form
dimers with their adjacent inner Na atoms as shown A or C in
fig.\ \ref{fig:distance}(a).
The same situation occurs for the three-caged cluster in 
fig.\ \ref{fig:distance}(b).
Here both the atoms forming the distances D and those forming 
the distance E try to form dimers. 
However they are balanced in force. 
Thus the length D is almost equal to that E\@. 

The distances between the endohedral Na atoms in hydrogen free four-caged
bamboo structure Si$_{78}$ are shown in fig.\ \ref{fig:distanceHfree}(a).
A single dimer exists at the center of the bamboo structure.
Since both the Na atom pairs at the edges have formed the edge state
forming the relaxed longer distance, the Na atoms just inside the bond
have formed the dimer. 
Fig.\ \ref{fig:distanceHfree}(b) shows the distances between the endohedral
Na atoms in the hydrogen free three-caged bamboo structure Si$_{60}$.
The inter-Na distances are the same for each endohedral atom;
a balance of forces exists between the central Na atom and the adjacent
two Na atoms.
These fig.\ \ref{fig:distance} and fig.\ \ref{fig:distanceHfree} indicate that
the Peierls distortion exists between the endohedral Na atoms in these
bamboo structures.

\begin{figure}
\begin{center}
\includegraphics{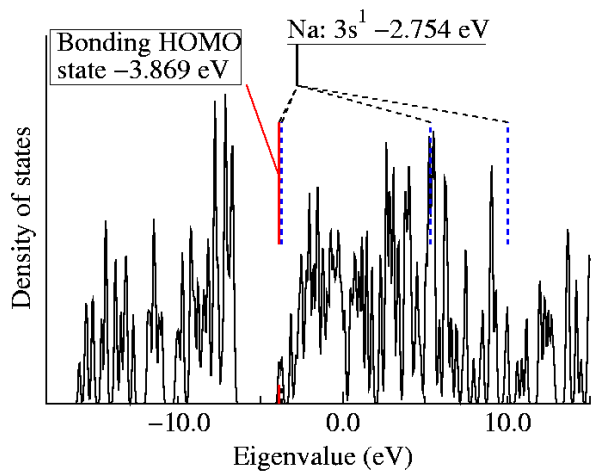}
\end{center}
\caption{\label{fig:dos} Molecular density of states (DOS)
of the double-caged Na$_2$@Si$_{42}$H$_{36}$ cluster.
The 3$s$ state of the isolate Na atom splits into the
bonding HOMO state and the several unoccupied anti-bonding states.}
\end{figure}

Fig.\ \ref{fig:dos} shows the molecular DOS of the double caged
Na$_2$@Si$_{42}$H$_{36}$ cluster. 
The shape of the earlier density of states~\cite{bib:Dong1999,bib:Madsen2003} of
the clathrate are similar to this density of states. 
The HOMO state is at $-3.869$ eV and the LUMO is $-3.703$ eV, where
HOMO is the highest occupied state and LUMO is the lowest unoccupied state.
The HOMO-LUMO
gap is $0.166$ eV. 
The magnitude of the LDA gap has been $0.177$ eV with the electron
correlation by Vosko, Wilk, and Nusair~\cite{bib:Vosko1980}.
No experimental band gap energy has been given, since HOMO is located
at just above the gap. 
The eigenvalue $-2.754$ eV for 3$s$ state of the isolate Na atom
splits into an occupied single bonding state $-3.869$ eV
(18A1.g) which is the HOMO state and several higher anti-bonding
states 16B3.u (LUMO, $-3.703$ eV), 28A1.g ($5.328$ eV) and 26B3.u
($10.02$ eV). 
The decrease of the eigenvalue from the 3$s$ at $-2.754$ eV to
the HOMO edge at $-3.869$ eV is due to the formation of the bonding
state between the endohedral atoms.
This is just the bonding state formation in hydrogen molecule.
Since the HOMO 18A1.g state is composed of a \textquotedblleft 
gerade\textquotedblright \ function, the corresponding electron state
gives even function with respective to the center of the molecule.
There is a large forbidden region from the HOMO down to the state 9A1.u
at $-6.593$ eV indicating the cluster is an insulator with
HOMO-LUMO gap $2.724$ eV, if the double caged cluster has no endohedral
atom.
For the four caged bamboo structure, the HOMO-LUMO gap has been $0.255$ eV.
This corresponds to a Peierls gap of this cluster. 

\begin{figure}
\begin{center}
\includegraphics{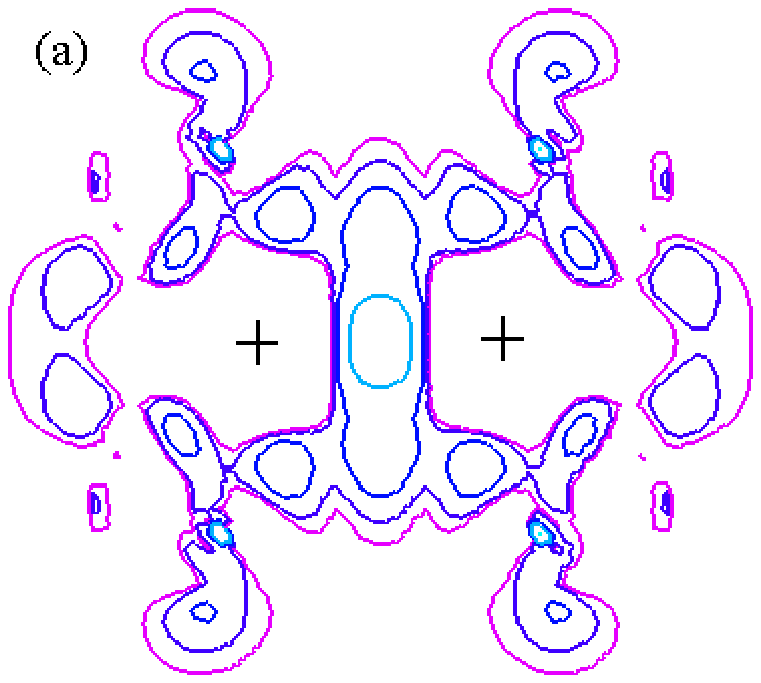}\\ 
\includegraphics{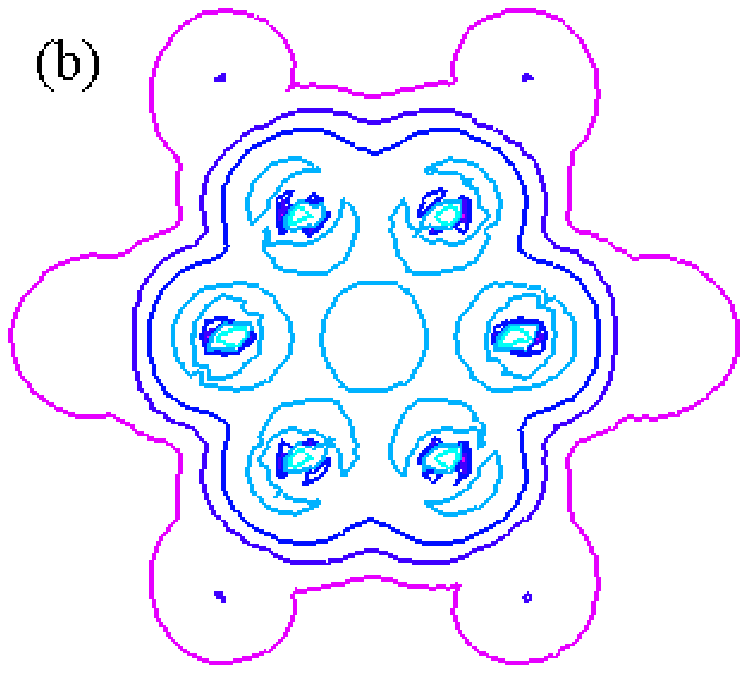}
\end{center}
\caption{\label{fig:4terms}
(Colour print) The spin unrestricted difference electron charge density profiles
$\Delta \rho _{\rm Na-Na}$ given by eq.\ (\ref{eq:4terms}), 
where the densities are plotted on a logarithmic scale, 
$10^{-5} \times 10^{5N/10}$ $e$/\AA $^3$, $N=0$-$10$. 
The blue lines are higher densities than the purple ones.
The two plus marks correspond to the positions of the endohedral Na atoms.
The blank regions correspond to the densities to be negative or less
than $10^{-5}$ $e$/\AA $^3$.
The density (a) is shown on the plane that intersects the two
endohedral Na atoms and the midpoint between two Si atoms on the
hexagonal ring shared by the adjacent two Si$_{24}$ cages.
The density (b) on the hexagonal ring between the two
adjacent Na atoms in the Si$_{24}$ cages.}
\end{figure}

To examine the bonding electron distribution between the endohedral Na
atoms in the double caged cluster, we show the electron density profile in
fig.\ \ref{fig:4terms} given by
\begin{eqnarray}
 \Delta \rho _{\rm Na-Na} &=&
 \left( \ {\rm Na} \ {\rm Na} \ \right)
+\left( \ \ \circ \ \ \circ \ \ \right) \nonumber \\
&-&\left( \ {\rm Na} \ \ \circ \ \ \right)
-\left( \ \ \circ \ \ {\rm Na} \ \right) \nonumber \\
 &=&
     \rho ({\rm Na}_2@{\rm Si}_{42}{\rm H}_{36})+\rho ({\rm Si}_{42}{\rm H}_{36}) \nonumber \\
     &-& \rho ({\rm Na}@{\rm Si}_{42}{\rm H}_{36})-\rho ({\rm Na}@{\rm Si}_{42}{\rm H}_{36}),
  \label{eq:4terms} 
\end{eqnarray}
where open circles represent the vacancies of the endohedral Na atoms. 
The coordinates of the last three terms are fixed at those of the first
term to obtain the difference of the electron charge densities. 
This expression gives the interaction electron density between the Na atoms,
since the net number of atoms is cancelled.
We have used the spin polarized calculations for all the the terms;
the non-spin states have been the lowest for the first two terms and
the spin states with $\mu _B=1$ have been the lowest for the last two terms.
We have evaluated the sum of the up-spin density and the down-spin density
for each structure and substituted them into the above equation and show
the density distribution in fig.\ \ref{fig:4terms}(a).
This figure shows a clear covalent bonding density
between the Na-Na bond.
This is formed by the dimer formation.
The density is due to the bonding state between each 3$s^1$ valence 
electron in the two Na atoms; this is just like the covalent 
bond formation between two hydrogen atoms.
We show in fig.\ \ref{fig:4terms}(b) the difference density on the
hexagonal ring located at the bisector plane between the two plus marks
in (a).
There is the finite covalent charge densities on the plane.
Neither the total electron density nor the partial charge density due
to the HOMO state in fig.\ \ref{fig:dos} has shown this type of the
covalent bond charge densities between the Na atoms.
The densities of the bonding states have appeared between the
dimers in the even number of cages.

\begin{figure}
\begin{center}
\includegraphics{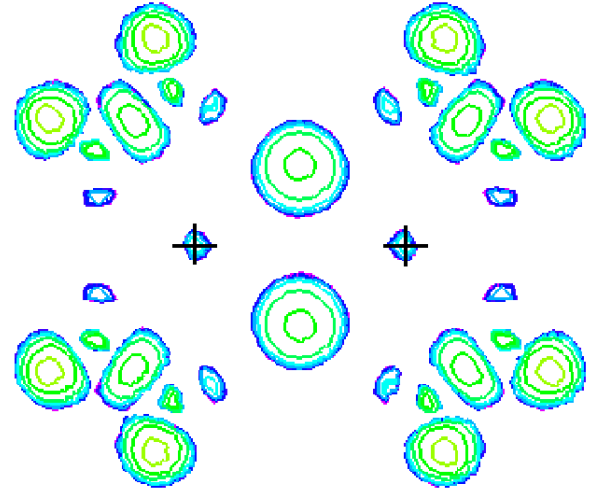}
\end{center}
\caption{\label{fig:2terms}(Colour print) The difference charge density
between the converged self-consistent electron and the overlapped
isolated atom density. 
The densities are plotted on the same plane as in fig.\ \ref{fig:4terms}(a)
with a logarithmic scale,
$10^{-5} \times 10^{5N/10}$ $e$/\AA $^3$, $N=0$-$10$. 
The green contours are higher than the purple ones.
The blank regions are lower area than $10^{-5}$ $e$/\AA $^3$
including negative densities.}
\end{figure}

To see the spatial distribution of the electron transfers around
the endohedral atoms, we show in fig.\ \ref{fig:2terms} the difference
electron density profile 
\begin{eqnarray} \label{eq:2terms}
\Delta \rho = \rho _{\rm opt} - \sum \rho _{\rm atom},
\end{eqnarray}
where $\rho _{\rm opt}$ is the density of the geometrically optimized
cluster and $\rho _{\rm atom}$ is the overlapped 
density of the isolated constituent atoms.
The blank zone corresponds to regions with lower densities than
$10^{-5}$ $e$/\AA $^3$ or with negative densities.
So the contour lines correspond to the increased charge ones
comparing with the overlapped isolated atom densities.
The electrons around the endohedral Na atoms are depleted to the
cage silicon atoms except for the nucleus positions of the Na atoms.

We calculate electron transfers from the endohedral Na atom
to the frame atoms. 
There have been several methods to calculate the transfer.
Among them the Mulliken charges have been found to depend on the number
of the linear combination of atomic orbitals for the basis
functions~\cite{bib:Guerra2003}. 
Voronoi charges, which have been named as Voronoi deformation charge VDC, 
have been found to gives reasonable values for the
transfer~\cite{bib:Guerra2003}.
The transferred 3$s^1$ electron from each endohedral Na
atom to the frame silicon atoms have been 0.320$e$ for the double caged
bamboo structure.
The ionic states also appeared in the triple caged bamboo structure:
The electron transfers from Na atoms to frame atoms have been 
0.343$e$ (middle Na atom) and 0.297$e$ (edge Na atoms) for the triple caged
bamboo cluster showing that the remained 3$s^1$ electron of the endohedral Na
atoms has formed the covalent bonding states between the endohedral Na
atoms as shown in fig.\ \ref{fig:4terms}.

Here we evaluate the cohesion energy of the Na chain in the clathrate I\@.
For this purpose we evaluate the energy using the periodic DFT method with
the same type equation as eq.\ (\ref{eq:4terms});
\begin{eqnarray} \label{eq:eb4terms}
E^c &=&
\begin{picture}(50,50)(0,22)
 \put(0,  0){\framebox(50,50){\ }}
 \put(0, 11){\makebox(25,25){Na}}
 \put(25,11){\makebox(25,25){Na}}
\end{picture}
+
\begin{picture}(50,50)(0,22)
 \put(0,  0){\framebox(50,50){\ }}
\end{picture}  \nonumber \\
&-&
\begin{picture}(50,50)(0,22)
 \put(0,  0){\framebox(50,50){\ }}
 \put(0, 11){\makebox(25,25){Na}}
\end{picture} 
-
\begin{picture}(50,50)(0,22)
 \put(0,  0){\framebox(50,50){\ }}
 \put(25,11){\makebox(25,25){Na}}
\end{picture} \nonumber \\
   & & \  \nonumber \\
   & & \  \nonumber \\
   &=& E_T({\rm Na}_2@{\rm Si}_{46})
      +E_T({\rm Si}_{46}) \nonumber \\ 
      &-&E_T({\rm Na}@{\rm Si}_{46})
      -E_T({\rm Na}@{\rm Si}_{46}),
\end{eqnarray}
where $E_T$'s are the total energies of each crystal.
The net number of each kind of atoms is also cancelled in this equation.
The last two terms correspond to that each Na atom is located at
infinitely separated positions in the clathrate.
Therefore this equation enable us to evaluate the cohesive energy of the Na
chain in the clathrate I\@.
The equation has been derived from the difference of the formation energies
of each phase by Sawada {\it et al}~\cite{bib:Sawada1140}.
They proposed this equation to evaluate the binding energies between the
substitutional solute atom and the interstitial solute atom in the bcc iron.
They needed to calculate each energy in supercells as large as possible. 
For our calculation the use of the unit cell is sufficient, since we need
to calculate the cohesion energy of the Na chain in the clathrate. 
We calculate four kinds of the equation of states for each clathrate
in eq.\ (\ref{eq:eb4terms}).
Here we have assumed the energies of the last two terms to be equivalent
due to their symmetry.
The equation used is
\begin{eqnarray} \label{eq:eb3terms}
E^c =E_T({\rm Na}_2@{\rm Si}_{46})
    +E_T({\rm Si}_{46}) 
    -2E_T({\rm Na}@{\rm Si}_{46}).
\end{eqnarray}
The energy of the first term have been the lowest for a spin-polarized
state $\mu _B=0.188$ and the other terms for the non-spin states. 
The equilibrium lattice constant of this clathrate has been
$10.1998$ \AA , the shorter inter-Na distance has been $5.0915$ \AA ,
and the longer one $5.1083$ \AA , showing the difference by 
$0.0168$ \AA \ in the [100] direction.
The difference between these two distances is smaller than that in the
hydrogenated cluster in fig.\ \ref{fig:distance}(a).
This is because of the infinite chain connection of the Na atoms in the 
clathrate I\@.
The cohesive energy $E^c$ of the chain has been $0.104$ eV which is
finite and attractive, so the chain is more stable than the two
infinitely separated Na atoms in the clathrate I\@.

To evaluate the energy gain of the distortion in the crystalline state,
we calculate the total energy of the clathrate in which 
the two Na atoms are located at the centers of the gravity of the
nearest Si$_{24}$ cages of the first term in eq.\ (\ref{eq:eb3terms}). 
This energy has been higher by 0.00186 eV, with the shorter inter-Na
distance $5.0978$ \AA , than the full relaxed clathrate.
The shorter inter-Na distance in the full relaxed clathrate is
shorter by $0.0062$ \AA \ than the inter-gravity distance.
This quantity is a significant difference in the accuracy of the DFT
calculations.
This also indicates the existence of the attractive interaction between
the shorter Na atom pairs.

\section{Discussion}
The endohedral atoms have interacted with the cage atoms through the
ionic bond and with the nearest endohedral atoms through the covalent
bond.

We have assumed that the clathrate I consists of the bamboo structures
in the three perpendicular directions.
Here, we examine the validity of this assumption.
We have calculated the total energy of the clathrate Na$_2$@Si$_{46}$ in
which one of the two Na atoms is located in Si$_{20}$ cage and the other
in Si$_{24}$ cage.
We have already calculated the energy of the clathrate Na$_2$@Si$_{46}$
in which two Na atoms are located at the nearest Si$_{24}$ cages.
The energy has been given as the first term in eq.\ (\ref{eq:eb3terms}).
The energy of this clathrate has been more stable by 0.189 eV than that
of the former clathrate;
the binding between the two Na atoms in the chain is more stable than
the two Na atoms in the Si$_{20}$ and Si$_{24}$ cages. 
This is another evidence of the validity of our presumption for the
structure of the clathrate I\@.  

The covalent bond charge has existed in fig.\ \ref{fig:4terms}
between the endohedral Na atoms.
The validity of our bamboo structure model for the clathrate I
is supported by experimental evidence of the preferential
occupation of the Ba atoms in the Si$_{24}$ cages by Yamanaka
{\it et al}~\cite{bib:Yamanaka103}. 
They reported that the Ba atoms occupy 0.985 of the six Si$_{24}$
cages and only occupy 0.189 of the two Si$_{20}$ cages. 
The high occupancy is a proof of the existence of the covalent
bond between the Ba atoms Si$_{24}$ cages.
No explanation has ever been given for the origin of the occupancies.

The present study has predicted the difference of the inter-Na
distances to be only $0.0168$ \AA \ in the [100] direction. 
No report has been existed for the experimental guest
displacement in the clathrate I except for the guest displacement
parameters~\cite{bib:Paschen2001,bib:Christensen2003}. 
This is because the displacement is too small to be measured.

The anisotropy of the atomic displacement parameters of the
endohedral atoms in clathrate I has been reported by
Chakoumakos \textit{et al} ~\cite{bib:Chakoumakos80,bib:Chakoumakos127},
Nolas \textit{et al}~\cite{bib:Nolas3845} in which
much smaller amplitudes in the [100] directions were reported than in
its perpendicular directions.
The present study explains the anisotropy to be due to the constrain 
of the displacements of the Na atoms in the [100] directions induced
by the covalent bond:
the bond constrains the displacements between the nearest Na atoms in
the directions.
No explanation has been given for the origin of the anisotropies. 

The covalent bond between the endohedral Na atoms prevents the
atoms from crystallizing into ordered ionic structure like NaCl or
CsCl and crystallizes into the caged clathrate structures.
The bond forms beads of the Na atoms in the clathrate
I or three dimensional network of the Na atoms with $T{_d}$ symmetry
in the clathrate II\@.
This is because the electro-negativity of the host 14 group atoms is
smaller than that of the halogen atoms that crystallize into ionic
crystals.
The smaller electro-negativity differences between the host and guest
atoms allow the guest Na atoms to form both the covalent bond between the
guest atoms and the ionic bond through the charge transfer to the cages.
Thus the clathrates are a compromised electronic state between the ionic
crystals and the covalent crystals.

\section{Conclusions}
Presuming that the clathrate I consists of the sheaf of one-dimensional
connections of Na@Si$_{24}$ cages interleaved in the three perpendicular
directions, we have investigated the binding nature of the endohedral Na
atoms with both the real-space and the periodic DFT methods.
Each Na atom has lost 30\% of the 3$s^1$ charge to the frame.
The finite covalent bonding charge due to the Peierls distortion has
existed between the endohedral Na atoms in the caged clusters. 
The cohesion energy has been $0.104$ eV for the chain in the [100]
directions of the clathrate I\@.
The presumption has been proved to be valid;
the clathrate encapsulating two Na atoms in the [100] direction has been
more stable by $0.189$ eV than the clathrate encapsulating the two atoms
in the Si$_{20}$ and Si$_{24}$ cages.
This covalent bond has explained the experimental anisotropic displacement
parameters and the preferential occupancies of the endohedral atoms
in the Si$_{24}$ cages of the clathrates I\@.
The difference between the Na-Na distances has been $0.0168$ \AA .
This small magnitude of the displacement coincides with the absence of
the experimental reports on the guest displacements in the clathrate I\@.
The beads of the endohedral Na atoms in the directions are due 
to the the covalent bond between the endohedral atoms accompanying 
the electron charge transfer from the endohedral atoms to the cages.
The covalent bond has explained both the preferential occupancies of the
endohedral atoms and the low anisotropic displacement parameters in the
[100] directions in the Si$_{24}$ cages of the clathrate I.
The beads are just a precipitated state in the regular solution theory.
The smaller electro-negativity of group 14 host atoms than the halogen
atoms allows the endohedral Na atoms to prevent the atoms crystallizing
into the ionic crystals and allows the atoms to form the covalent bonds
with beads of the endohedral atoms.

\begin{acknowledgments}
Computations were performed in part using SCore systems at the
Information Science Center in Meiji University and Altix 3700 BX2 at
YITP in Kyoto University.
\end{acknowledgments}

\bibliography{tomono}

\end{document}